\documentclass[epj]{svjour}
\usepackage{graphicx}
\usepackage{bm}
\usepackage[english]{babel}
\usepackage{amsmath,times}
\usepackage{amsfonts}
\usepackage{ulem}
\usepackage{soul}
\usepackage{color}
\usepackage{amssymb}

\def\ket#1{\mathinner{|{#1}\rangle}}

\newcommand{\mgap}[1]{{\color{black}#1}}
\begin{document}
\title{Quantum thermometry by single-qubit dephasing}
\author{Sholeh Razavian\inst{1} \and Claudia Benedetti\inst{2} \and
Matteo Bina \inst{2} \and Yahya Akbari-Kourbolagh \inst{1}
\and Matteo G. A. Paris \inst{2}}
\institute{Faculty of Physics, Azarbaijan Shahid Madani University, Tabriz, Iran 
\and Quantum Technology Lab, Dipartimento di Fisica {\it Aldo Pontremoli}, 
Universit\`a degli Studi di Milano, I-20133 Milano, Italy}
\date{Received: date / Revised version: date}
\abstract{
We address the dephasing dynamics of a qubit as an effective 
process to estimate the temperature of its environment. Our scheme 
is inherently quantum, since it exploits the sensitivity of the qubit 
to decoherence, and does not require thermalization with 
the system under investigation.  We optimize the quantum Fisher 
information with respect to the interaction time and the temperature in the case of Ohmic-like environments. 
We also find explicitly the qubit measurement achieving the quantum Cram\'er-Rao 
bound to precision. Our results show that the conditions for optimal 
estimation originate from a non-trivial interplay between the dephasing 
dynamics and the Ohmic structure of the environment. In general, optimal 
estimation is achieved neither when the qubit approaches the stationary 
state, nor for full dephasing.} 
\PACS{XXX} 
\maketitle
\section{\label{sec:level1}INTRODUCTION}
Thermometry is the art of inferring the temperature of a (large) 
sample by reading its value on a (smaller) probe. Standard thermometry 
is based on the zero-th law of thermodynamics: the sample is assumed 
to have a large heat capacity whereas the probe (i.e. the thermometer) 
has a much smaller one. They are put in contact, and after a while 
they achieve thermalization by exchanging energy. The sample has not changed its temperature, owing to its large heat capacity, whereas the
thermometer is now at the same temperature of the sample, which 
may be measured by the experimenter. In practice, since the heat capacity of 
the sample is always finite, a temperature measurement always implies 
a disturbance of the temperature itself. Besides, for any quantum 
system with a non-vanishing gap, precise thermometry cannot be 
achieved below a certain threshold temperature \cite{parislandau,hofer}.
\par
Quantum probes, exploited as quantum thermometers, may offer 
a different and more effective avenue to thermometry. In turn, 
the use of quantum probes and quantum measurements for thermometry has 
attracted much attention \cite{Qthermo1,Qthermo2}, with coherence 
and interference effects playing a relevant role 
\cite{stace,stadp,adesso,jevtic,johonson,geno18,Steinberg,Barbieri}.
Quantum probes offers the possibility of sensing temperature with small
disturbance \cite{brunelli,brunelli2}, i.e leaving the thermal system mostly unperturbed. In this framework, single-qubit probes are perhaps 
the best choice, being the simplest and smallest system suitable 
to extract information from the sample \cite{Qthrmo,horodecki}. 
Besides, it is of interest to exploit estimation scheme not based on 
the zero-th law of thermodynamics, i.e. on the exchange of energy between 
the sample and the thermometer \cite{equilib,Kroner}.
 In this paper, we address in details the dephasing dynamics of a qubit as an effective process to estimate the temperature of its environment. This kind of scheme is inherently quantum, since it exploits the fragility and the sensitivity of quantum systems to decoherence, and does not require thermalization between the qubit probe and the system under investigation. 
\par
Indeed, the temperature of a quantum system is not a 
quantum observable. In other words, in a quantum setting 
temperature maintains its thermodynamical meaning, but it loses 
its operational definition. Therefore, any strategy 
aimed to determine temperature ultimately reduces to a 
parameter estimation problem, more precisely to a 
{\it quantum parameter estimation} problem \cite{paris}. 
The scope of quantum estimation is to provide 
an estimate of the unknown parameter from repeated measurements 
on the probe. The experimenter chooses a quantum measurement 
on the probe and then process the data. The choice of an estimator 
corresponds to a classical post processing of the outcomes after 
the measurement, whereas the choice of the measurement is the
central problem of quantum metrology, since different measurements 
lead to different precisions. Indeed, quantum parameter estimation has been successfully applied to a variety of topics, ranging from phase estimation \cite{Holevo,Dariano,BinaOlivares,Monras} and open system dynamics \cite{Milburn,Rossi}, to quantum measurements in qubit chains \cite{Tamascelli,Burgarth} and quantum phase transitions \cite{Zanardi,Amelio,RossiBina}.
\par
In order to achieve optimal quantum thermometry, i.e. an 
estimator with minimum fluctuations, we will 
use notions and tools from quantum estimation theory. Notice, 
however, that we are not discussing here fluctuations of 
temperature in the thermodynamical sense. As a matter of fact, 
the temperature 
of the sample itself may not fluctuate \cite{Qthrmo}. On the other 
hand, the inferred value of temperature, i.e. the temperature 
estimate extracted from measurements  performed on the qubit, 
does indeed fluctuate \cite{web06,jah11}.
\par
Concerning the interaction model, we assume that the interaction 
of the qubit with the sample is described by a dephasing Hamiltonian. 
The model is exactly solvable \cite{breuer,palma96} when 
the environment (i.e. the sample under investigation) is excited in 
a thermal state. We let the qubit interact with 
its environment, and then we perform a measurement in order to extract 
information about the temperature. Our scheme is valid for a generic
sample without any restriction on its energy spectrum. However, for the sake
of providing some quantitative results, we assume an Ohmic spectral 
density with a generic Ohmicity parameter \cite{Paavola,Martinazzo,Myatt,PiiloManiscalco,Grasselli,claudia}. 
The interaction time is a free
parameter, which we employ to maximize the qubit quantum Fisher information, 
i.e. the information about the temperature encoded in the state of the
qubit as a result of the interaction with the sample. As we will see,
the optimal interaction time is finite, i.e. the qubit is not required
to approach its stationary state, nor it corresponds to a full dephasing.
Rather, it is determined by a non-trivial interplay between the dephasing 
dynamics and the Ohmic structure of the environment \cite{a1,a2,a3,a4}, 
especially at low temperature.
\par
The paper is structured as follows. In Section \ref{sec:2}, we briefly 
review the tools of local quantum estimation theory (QET), 
whereas in Section
\ref{sec:3} we describe in some details our physical model, and how 
we exploit QET techniques in our system. In Section \ref{sec:4} 
we illustrate our results and show how to achieve optimal estimation with
feasible measurements. Section \ref{sec:5} closes the paper with 
some concluding remarks.
\section{\label{sec:2} Local quantum estimation theory}
A parameter estimation scheme is a procedure in which a 
quantity of interest, say the parameter $T$, is not 
measured directly, but rather inferred by processing the 
data from the measurement of a different observable, say $X$, 
which may directly be measured on the system under investigation.
We denote by $p(x|T)$ the conditional distribution of the 
outcomes of $X$, when the true value of the parameter is $T$,
and by $\hat T ({\mathbf x})$, ${\mathbf x} =\{x_1, x_2, ..., x_M\}$ 
any {\it estimator}, i.e. 
a function mapping the $M$ observed outcomes to a value of the 
parameter. The estimated value of the parameter is the average
value of the estimator 
\begin{equation}
\bar T = \int\!\! d{\mathbf x}\, p({\mathbf x}|T)\,
\hat T ({\mathbf x})\,,
\end{equation}
whereas the overall precision of the estimation procedure is quantified
by its variance
\begin{equation}
\hbox{Var}\, \hat T = \int\!\! d{\mathbf x}\, p({\mathbf x}|T)\,
[\hat{T}({\mathbf x})-\bar T]^2\,.
\end{equation}
In both equations $p({\mathbf x}|T) = \Pi_{k=1}^M\, p(x_k|T)$ 
since measurements are performed on repeated preparations of 
the system and thus are independent.
As a matter of fact, the variance of any unbiased estimator 
(i.e. an estimator for which $\bar T \rightarrow T$ in the asymptotic
limit $M \gg1$) for the parameter $T$ is bounded by the Cram\'er-Rao 
inequality, stating that
\begin{equation}
\hbox{Var}\, \hat T \geq \frac{1}{M F(T)},
\end{equation}  
where $F(T)$ is the so-called Fisher information of the measurement 
of $X$, which is given by
\begin{equation}\label{classFI}
F(T) = \int\!\!d\mathbf x\, p(\mathbf x|T)\left[\partial_{T}\log p(\mathbf x|T)\right]^{2}\,,
\end{equation}
where $\partial_{T}$ denotes the derivative with respect to the parameter $T$.
The optimal measurement for the parameter $T$ is the measurement with
the largest Fisher information, whereas an {\it efficient} estimator
is an estimator saturating the Cram\'er-Rao inequality. The combination
of the optimal measurement with an efficient estimator provides an 
optimal estimation scheme for the parameter $T$.
\par
The maximization over all the possible quantum measurements may be indeed
performed. The corresponding Fisher information is usually referred to as the {\it Quantum Fisher Information} (QFI) $ H(T)$ \cite{Cramer,Helstrom,Braunstein}. The ultimate 
precision allowed by quantum mechanics is thus achieved by an 
estimator that saturates the quantum Cram\'er-Rao bound    
\begin{equation}
\hbox{Var}\, \hat T\geq\frac{1}{M H(T)}\,.
\end{equation}
The expression of the QFI can be obtained 
from the quantum state of the system, and in particular 
from its eigenstates and eigenvectors, which contain the 
dependence on the parameter. Starting from 
the diagonal form of the density matrix describing the state of the 
system 
\begin{equation}
\varrho_{T}=\sum_{n}\rho_{n}|\phi_{n}\rangle\langle\phi_{n}|\,,
\end{equation} 
the QFI is given by    
\begin{equation}
H(T)=\sum_{p}\frac{(\partial_{T}\rho_{p})^{2}}{\rho_{p}}\,+\, 2\sum_{n\neq m}\frac{(\rho_{n}-\rho_{m})^2}{\rho_{n}+\rho_{m}}|\langle\phi_{n}|\partial_{T}\phi_{m}\rangle|^{2}\,. \label{ht}
\end{equation}
The first term in Eq. (\ref{ht}) depends on how the eigenvalues 
of state depends on the parameter and it is referred to as the {\it classical} 
part of QFI, whereas the second term is referred to as the {\it quantum} part, and takes into account the dependence of eigenvectors on the parameter of interest. We point out that the local QET presented so far, dependent on the specific value of the parameter $T$, implicitly assumes a prior rough knowledge of the parameter of interest \cite{paris}.
\par
In order to quantify estimability of a parameter independently 
on its value, one may introduce the signal-to-noise ratio (SNR)
$R_T=T^2/\hbox{Var}\, \hat T$, which is larger for better estimators.
Upon using the quantum Cram\'er-Rao inequality we have the bound
\begin{equation}
R_{T}\leq Q_T \equiv T^{2}H(T)\,, \label{qsnr}
\end{equation}
where $Q_T$ is usually referred to as the quantum signal-to-noise ratio
(QSNR). The larger is QSNR the more effectively estimable is 
the parameter $T$.
\par
In the following, we will exploit the framework described above 
in order to estimate the temperature $T$ of a structured sample
with Ohmic spectral density. The estimation strategy involves a qubit 
interacting with the sample for a certain interaction time that is
then measured in order to infer the temperature.
In particular, we investigate whether an effective
\begin{figure*}[t]
	\center
	\includegraphics[width=0.6\textwidth]{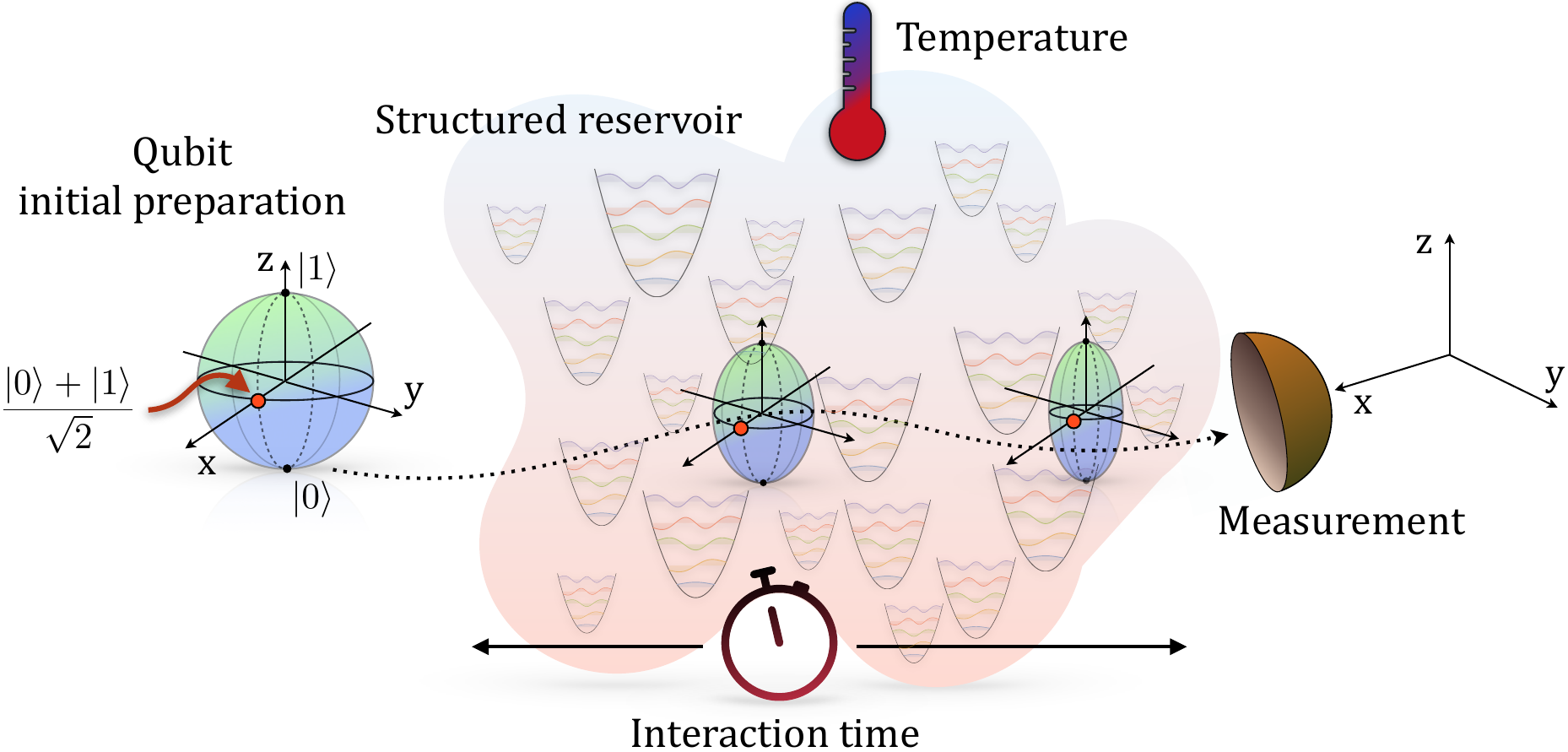} 
	\caption{(Color online) Quantum thermometric scheme based on 
		a single two-level system undergoing dephasing due to the 
		interaction  with a structured reservoir at thermal equilibrium.
		The dephasing mechanism is represented as a shrinking of the Bloch 
		sphere (in interaction picture). After interaction a measurement 
		of the optimal spin direction is performed.
	} \label{probscheme}
\end{figure*}
 quantum 
thermometric scheme may be achieved by considering a 
(exactly solvable) dephasing interaction model for the qubit. 
To this aim, we evaluate the optimal interaction time to achieve 
the maximum QFI and compute the corresponding QSNR. As we will see, 
we have encouraging numerical results in all the considered cases, 
and also few analytic results in the super-Ohmic regime, for 
low temperature in Ohmic environment, and in the high 
temperature regime for all the spectral densities.     
\section{\label{sec:3}QUANTUM THERMOMETRY BY DEPHASING}
Any interaction between a quantum system and its 
environment modifies the phases between 
the different components of its wave-function. This usually 
produces dephasing \cite{claudiaparis,addis} and, in turn, 
decoherence, due to the interaction among the system and the 
different modes of the thermal bath. This mechanism may be 
exploited to make the quantum system an effective 
probe to estimate parameters of the environment, without 
undermining the energy of the involved systems. In particular, 
in Fig.~\ref{probscheme} we 
illustrate a schematic diagram of our probing strategy for quantum 
thermometry by qubit dephasing. The dephasing mechanism is 
represented as a shrinking of the Bloch sphere, where the spin 
state (red dot) is in the interaction picture. After interacting with the sample 
(see discussion in Sec.~\ref{sec:4}) the probe is measured in the 
optimal spin direction.
\par
The qubit interacts with a structured reservoir at thermal equilibrium, 
characterized by a spectral density of Ohmic type.
The total Hamiltonian can be written as (we set $\hbar$ =1 and the Boltzmann constant  
$k_B$=1)
\begin{equation}
\mathcal{H}=\frac{1}{2}\omega_0\sigma_z + \sum_{k}\omega_{k}b_{k}^{\dagger}b_{k}+\sigma_{z}\sum_{k}(g_{k}b_{k}^{\dagger}+g_{k}^{\ast}b_{k})\ ,
\end{equation}
where  $ \omega_0 $ 
is the probe transition frequency between the ground state $\ket{0}$ and the excited state $\ket{1}$, $\omega_k $ are 
the frequencies of the reservoir modes, $ b_k(b_{k}^{\dagger})$ is
the bosonic annihilation (creation) operator for mode $ k $ and $ g_k $ are
the coupling constants of each mode with the qubit, which can be distributed according to different spectral densities.
\par
The qubit probe is initially prepared in a pure state 
$|\psi\rangle=\cos\frac{\theta}{2}|0\rangle + \sin\frac{\theta}{2}|1\rangle$ and the environment is supposed to be
in an equilibrium thermal state at temperature $T$, namely $\varrho_B= \exp\{- \mathcal{H}_B/T\}/\mathcal{Z}$, with $\mathcal{Z}=\hbox{Tr}[\exp\{- \mathcal{H}_B/T\}]$ the partition function and $\mathcal{H}_B$ the Hamiltonian of the bath.
Going to the interaction picture, the reduced open system dynamics 
of the probe is governed by the map
\begin{equation}\label{rht}
\varrho_S(t) = \hbox{Tr}_B \left[U_{I}(t)\,\varrho_{SB}(0)\,
U_{I}^{\dagger}(t)\right]\,,
\end{equation}
where $U_{I}(t)$ is the interaction-picture evolution operator and $\rho_{SB}(0) = |\psi\rangle\langle\psi| \otimes \varrho_B$ is the initial 
state of the whole system.
\par
Upon explicitly performing the trace over the degrees of freedom of the environment in Eq.~\eqref{rht} the 
evolved density matrix of the probe at time $t$ may be written as
\begin{equation}\label{rho}
\varrho_S(t)=\left(
\begin{array}{cc}
\cos^{2}{\frac{\theta}{2}} &\dfrac{1}{2} e^{-\Gamma(T,t)} \sin{\theta}\\
\dfrac{1}{2}e^{-\Gamma(T,t)} \sin{\theta} & \sin^{2}{\frac{\theta}{2}}
\end{array}
\right)\,,
\end{equation}
where the time-dependent behavior of the off-diagonal terms depends
on the decoherence factor
\begin{equation}
e^{-\Gamma(T,t)} =\sum_{k}\langle \exp(g_{k}b_{k}^{\dagger}-g_{k}^{*}b_{k})\rangle\, ,
\end{equation}
where $\langle \, \bullet \, \rangle = \hbox{Tr}[\bullet\,\varrho_B]$ 
is the average over the thermal state of the bath $\varrho_B$.
The decoherence factor $ \Gamma(T,t)$
depends on the temperature and on the spectral distribution 
of the coupling frequencies of the bath, which is 
quantified by the spectral density $J(\omega)$. The spectral density, 
in the continuous limit of the bath modes and for Ohmic-like distributions, 
is given by
\begin{equation}
J(\omega) \equiv J_s (\omega,\omega_c) = \frac{\omega^s}{\omega_c^{s-1}}\, e^{-\omega/\omega_c}\, ,
\end{equation}
where $ \omega_c $ is the cutoff frequency and $ s $ is the the Ohmicity parameter, which allows to classify the structured environment into three main classes, i.e. sub-Ohmic ($ s<1  $), Ohmic ( $ s=1$) and 
super-Ohmic ($ s>1 $).
The corresponding decoherence factors are, thus, given by
\begin{align}
\Gamma(T,t) = \int\!\! d\omega\, J (\omega)\, \frac{1-\cos \omega t}{\omega^2}\, \coth\left(\frac{\omega}{2T} \right)\label{gammaall}\,,
\end{align}
 where all the physical quantities in this equation should be 
 considered dimensionless and rescaled with the cutoff frequency $\omega_c$.
\mgap{
Upon expanding the hyperbolic cotangent as 
$\coth(x)=1+2\sum_{n=1}^\infty {\rm e}^{-2 n x}$ we may write
\begin{align}\label{gmtt}
\Gamma(T,t) = \Gamma( 0,t) + 2 \sum_{n=1}^\infty a_n^{1-s}\, \Gamma 
\left(0,\frac{t}{a_n}\right)\,, \qquad a_n \equiv a_n (T) = 1+ \frac{n}{T}\
\end{align}
where
\begin{align}
\Gamma (0,t) & =\int_0^{\infty}\!d\omega  \frac{1-\cos(\omega t)}
{\omega^2}\,J(\omega) =  \left( 1-\frac{\cos\left[(s-1)\arctan t\right]}{\left(1+t^2\right)^{\frac{s-1}{2}}}\right)\Gamma[s-1]
\label{gamtau}
\end{align}
$\Gamma[z]$ being the Euler Gamma function. Notice that for $s\rightarrow 1$
we have $\Gamma (0,t) \stackrel{s\rightarrow 1}{\longrightarrow}\frac12 \log\left(1+t^2\right)$.
Upon substituting this expression in Eq. (\ref{gmtt}), one realises that 
the series may be analytically evaluated, leading to
\begin{align}\label{gex}
\Gamma( T,t) = \Gamma (0,t) + s(s-1) T^{s-1} \frac{\Gamma[s-1]^2}{\Gamma[s+1]} 
\left[2\zeta(s-1,1+t) - \zeta(s-1,1+t-iT) -\zeta (1+T+iT)\right]
\end{align}
where $\zeta(p,z) = \sum_{n=0}^\infty (z+n)^{-p}$ is the generalised 
(Hurwitz) Zeta function.
Using this expression, one can see that the decoherence factor
is a monotonically increasing function of time $t$ for $s\leq 1$ whereas it
saturates to a constant value for $s>1$. \par
Given the decoherence factor, one may evaluate the QFI for 
the family of qubit states in Eq. \eqref{rho} \cite{claudia}, showing 
the optimal  initial preparation corresponds to  
$ \theta=\frac{\pi}{2} $.} The explicit expression of the QFI 
in this case  is given by 
\begin{align}
H(T,t) = \frac{\left[\partial_T \Gamma(T,t)\right]^2}{e^{2 \Gamma(T,t)}-1}\,.\label{fisher}
\end{align}
Thanks to the scaling properties of $\Gamma(T,t)$, the QFI, 
and in particular its optimized value at the optimal
interaction time $t_\text{opt}$, depends only on the Ohmicity parameter $s$ 
of the sample under investigation and on its temperature $T$.  
\mgap{Notice that using Eq. (\ref{gex}) an analytic expression may be 
derived for the QFI, which is however rather cumbersome and it will
not be reported here. }
\section{\label{sec:4}QUANTUM THERMOMETRY OF OHMIC-LIKE SAMPLES}
\begin{figure}[h!]
	\center
	\includegraphics[width=0.6\columnwidth]{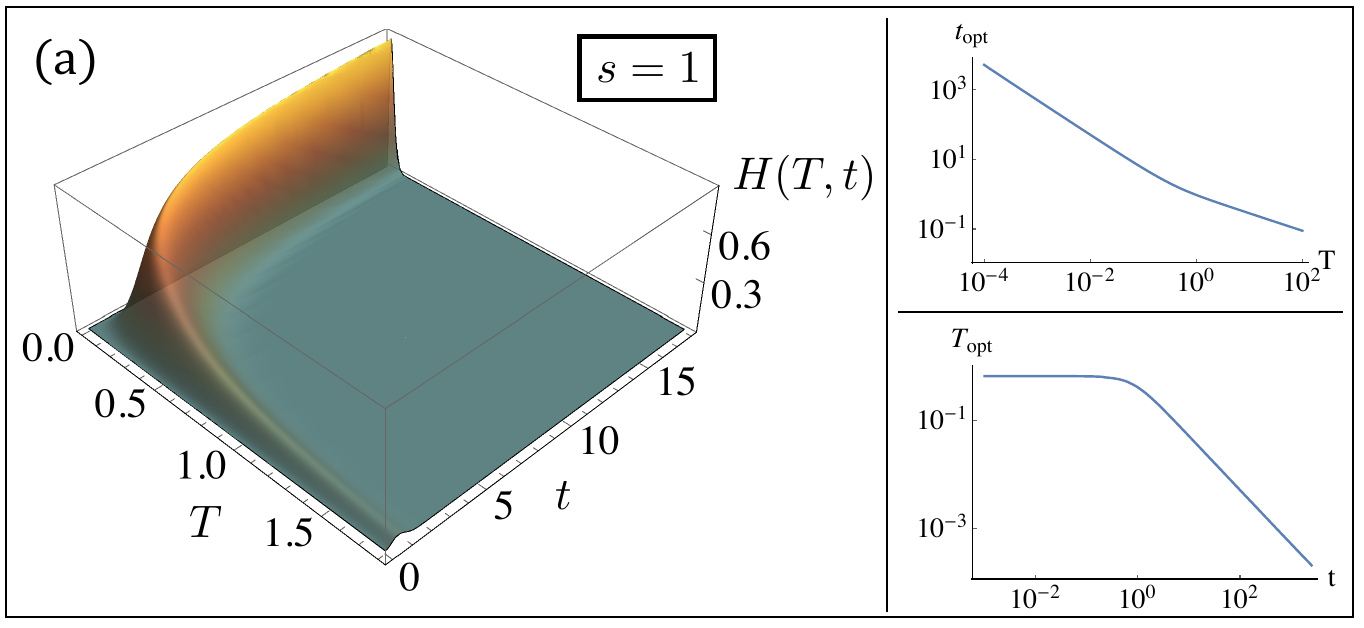}
	\includegraphics[width=0.6\columnwidth]{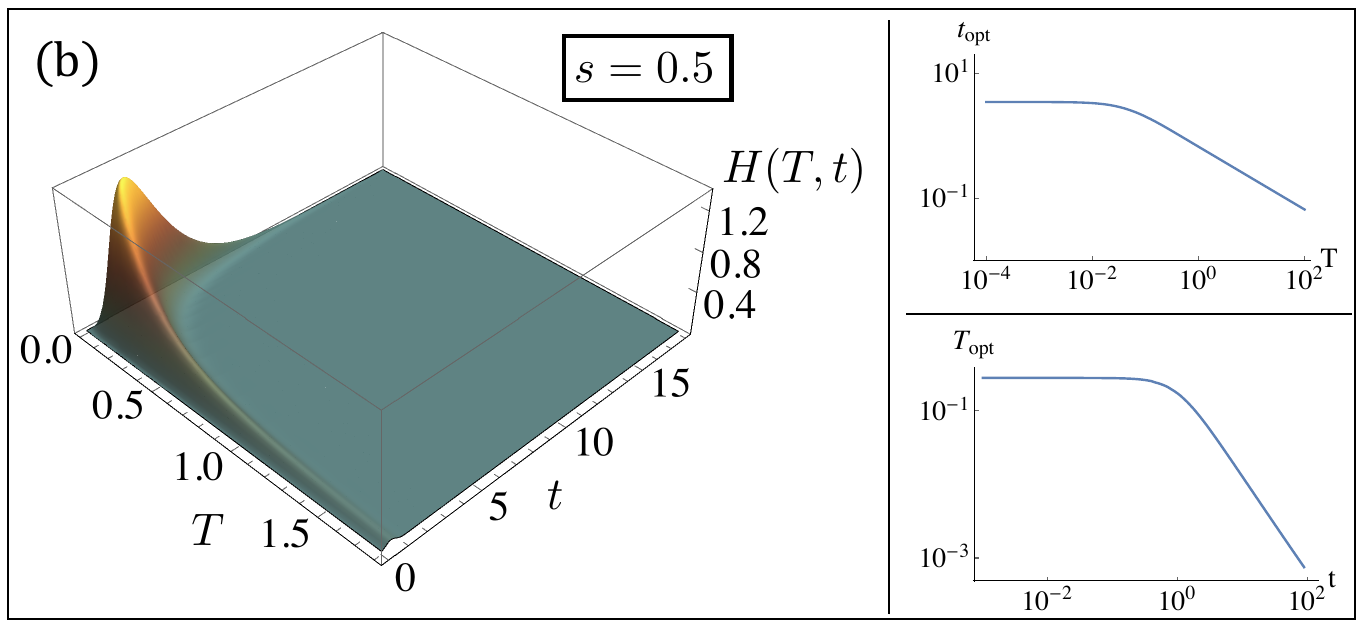}
	\includegraphics[width=0.6\columnwidth]{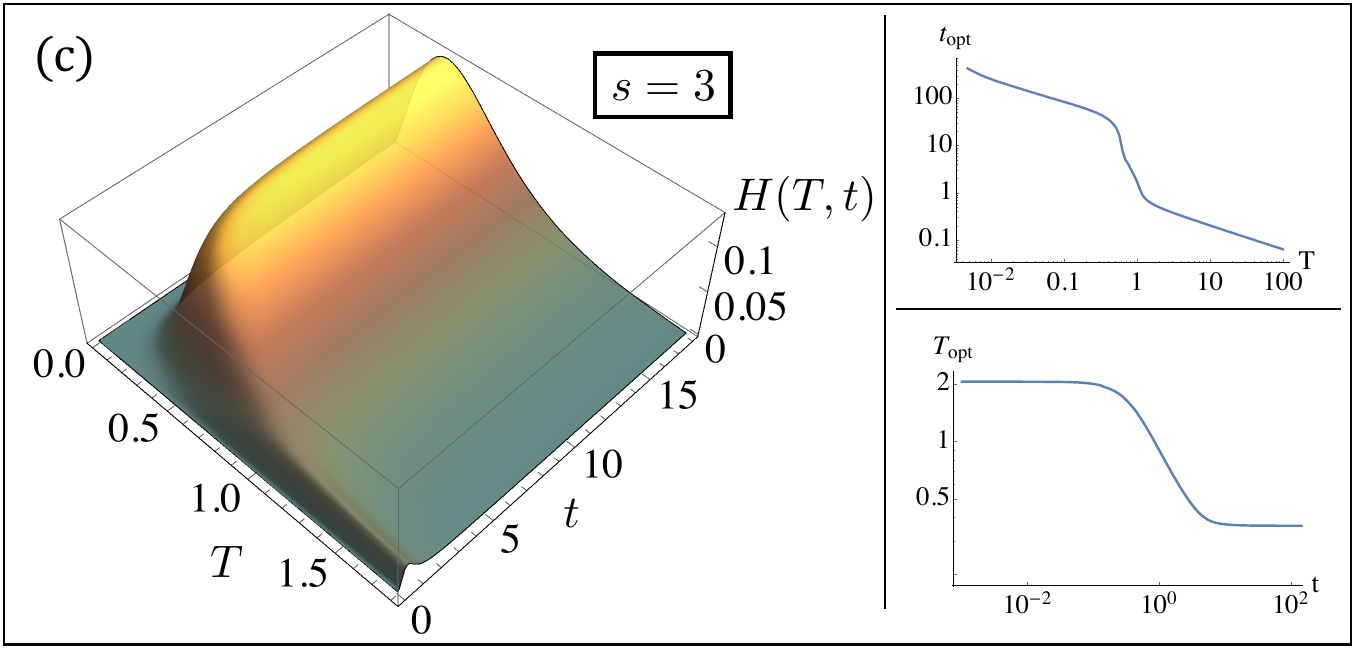} 
	\mgap{\caption{(Color online) The left boxes in each panel show plots 
	of the QFI $H(T,t)$ as a function of temperature and time
	for three illustrative cases of (a) Ohmic ($s=1$), (b) sub-Ohmic 
	($s=0.5$) and (c) super-Ohmic ($s=3$) environments. In the upper-right 
	boxes we plot the optimal interaction times $t_\text{opt}$ as a 
	function of reservoir temperature $T$. In the lower-right boxes, 
	we plot the temperature $T_\text{opt}$ at which the QFI is maximized, 
	as a function of the interaction time $t$.} \label{HtoptT}}
\end{figure}

In the following, we consider quantum thermometry of different 
samples having spectral density belonging to the Ohmic family. In particular, 
we consider three specific values of the Ohmicity parameter, 
corresponding to paradigmatic examples of sub-Ohmic ($s\!=\!0.5$), Ohmic ($s\!=\!1$) and super-Ohmic ($s=3$) environments. The behaviour of the QFI and 
the results of its maximization are shown in Fig.~\ref{HtoptT}. 
In the left boxes of each panel, we show a three-dimensional plot 
of $H(T,t)$, the QFI as a function of temperature $T$ and time $t$. 
As it  is apparent from the plots, at fixed temperature for $s=1 $ 
and $ s=0.5 $, there is a maximum for the QFI as
a function of time, whereas in the super-Ohmic case $s=3$ 
we see maxima for high temperature, whereas in the low temperature
regime we see saturation of the QFI, without a single local maximum
 value at a specific time. When a maximum for the QFI exists, it means 
that optimal estimation of temperature may be achieved at a finite
interaction time $t_\text{opt}$, i.e. before the qubit has reached its 
stationary state. 
\par
\par
\begin{figure}[h!]
	\center
	\includegraphics[width=0.5\textwidth]{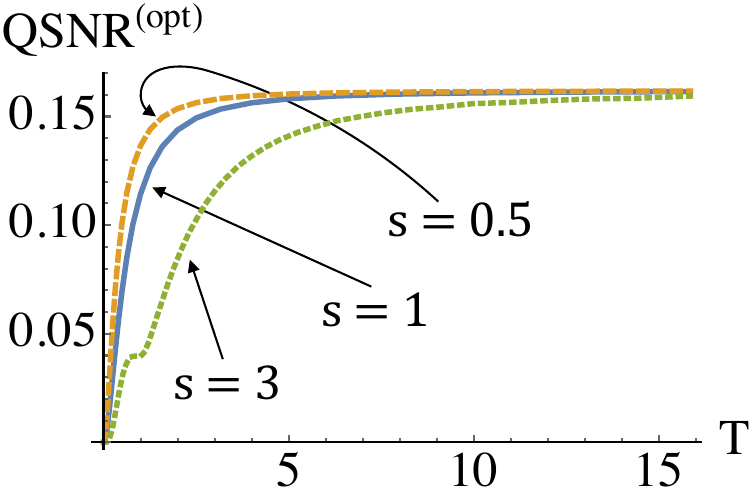} 
	\caption{(Color online) Plot of the QSNR evaluated at the optimal interaction time $t_\text{opt}$, as a function of the reservoir temperature $T$. We considered three classes of structured environments, Ohmic ($s=1$), sub-Ohmic ($s=0.5$) and super-Ohmic ($s=3$).  } \label{QSNR}
\end{figure}

The right boxes of each panel in Fig.~\ref{HtoptT} show two plots for 
each value of the Ohmicity parameter. In the upper ones we show 
the optimal interaction time $ t_\text{opt} $, i.e. the time at which 
the QFI reaches 
its maximum value, for a fixed value of  the reservoir temperature. 
The  different slopes in the plots identify different thermal 
regimes, say of {\it low} and {\it high} temperatures. 
\mgap{At low temperature the optimal time is larger, while for increasing
temperature it progressively decreases. This behaviour may have been 
intuitively expected given the nature of the probing technique. 
Indeed, the ability of the probe to extract information about
the temperature of the environment comes from its fragility 
against decoherence. At low temperature, decoherence is weaker
and it takes time to imprint information on the probe, while
at high temperature decoherence is faster. Notice also that at 
high temperature decoherence is mostly due to thermal fluctuations
and the structure of the environment is no longer relevant. This 
is reflected in the behaviour of the QFI at high temperature, which
is similar for the three values of $s$. On the contrary, the structure
of the environment is crucial to determine the decoherence mechanism
at low temperature, and this corresponds to a strongly $s$-dependent
behaviour of the QFI (see also the discussion below on the quantum signal-to-noise ratio and Fig.~\ref{QSNR}).}
The lower plots show the values $T_\text{opt}$, i.e. the  temperature 
maximizing the QFI at a given interaction time $t$. This is a relevant 
quantity to consider in the case of \textit{small samples}, where 
the interaction  time may be indeed limited, especially when a 
traveling qubit is employed as a quantum probe. 
\par
\begin{figure}[h]
\center
\includegraphics[width=0.55\textwidth]{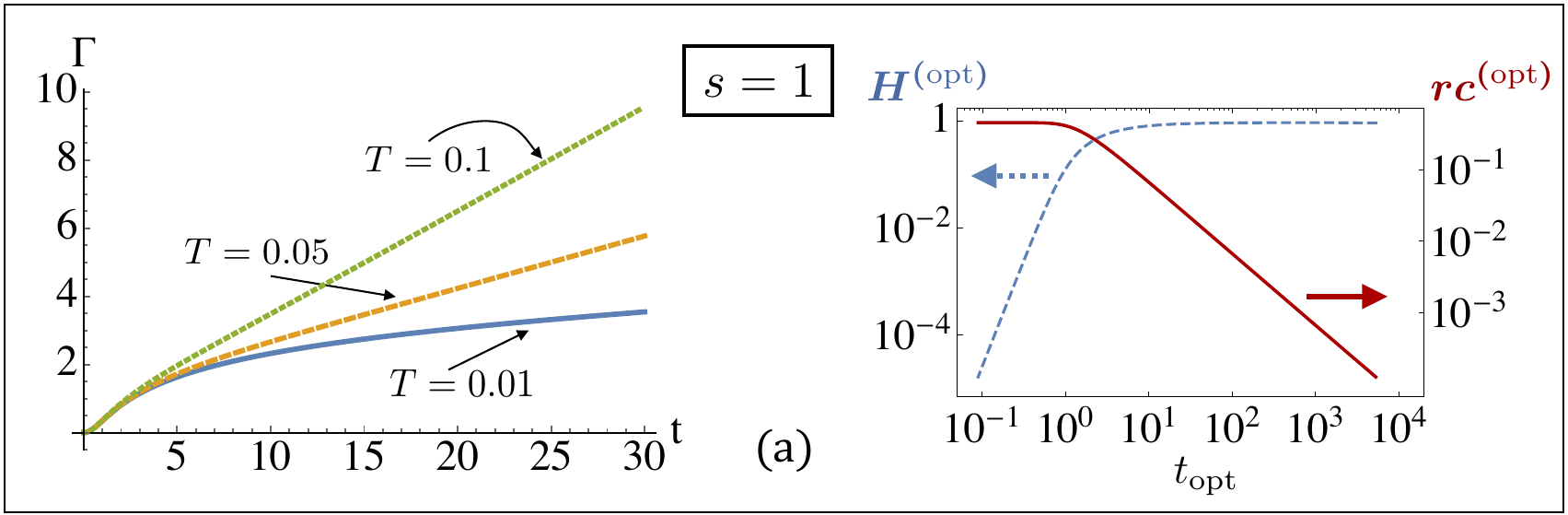} 
\includegraphics[width=0.55\textwidth]{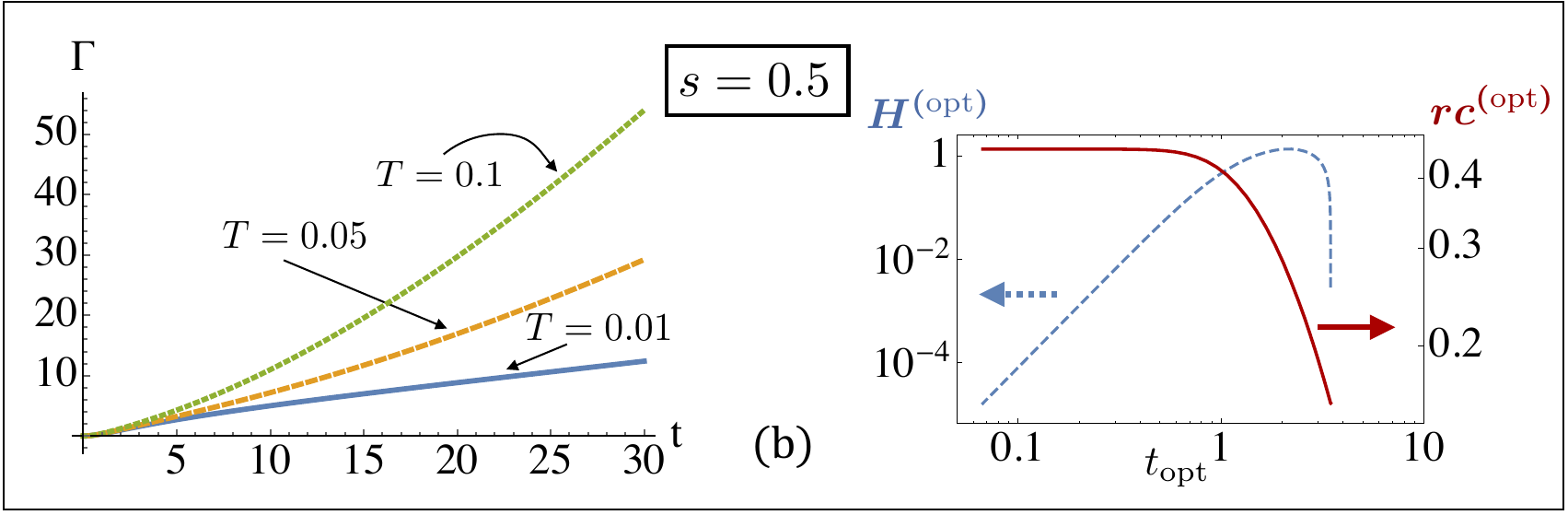} 
\includegraphics[width=0.55\textwidth]{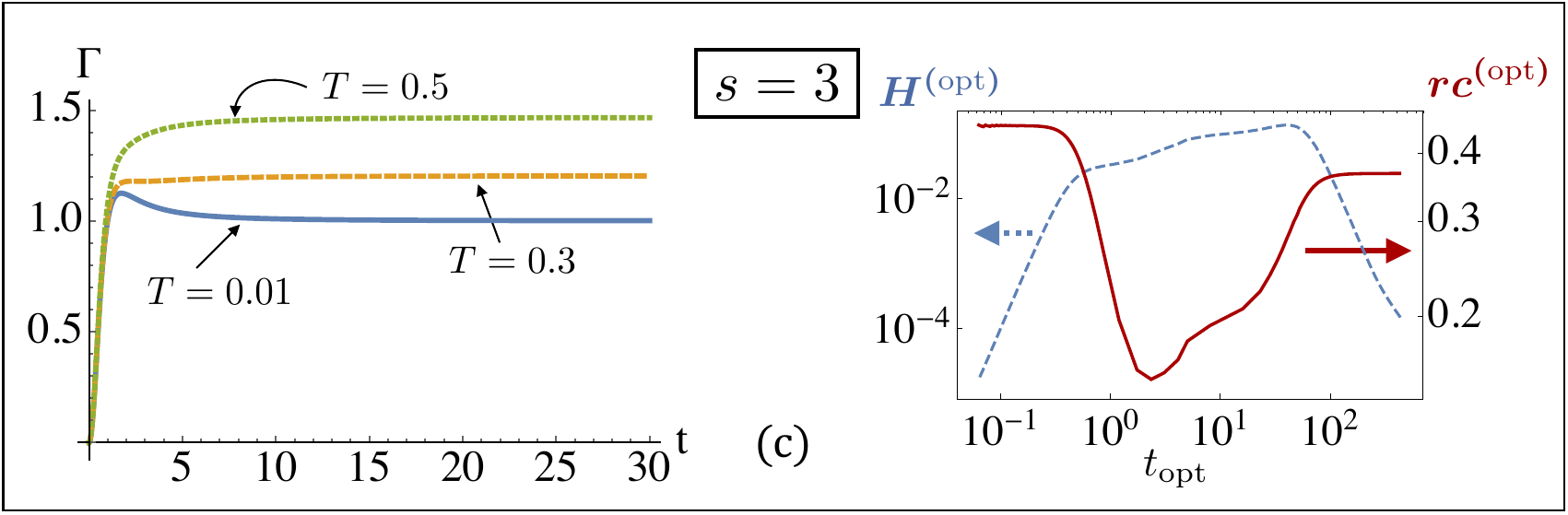} 
\caption{(Color online) On the left: plots of the decoherence factor $\Gamma(T,t)$ as a function of the interaction time $t$ for three values of the reservoir temperature (see solid, dashed and dotted curves). On the right: plots of the QFI $H^{(\text{opt})}$ (dashed curve) and the residual coherence $rc^{(\text{opt})}$ (solid curve), both evaluated at the optimal time $t_\text{opt}$. We notice that in order to reach high values of the QFI the probe must interact enough time with the reservoir, which, inevitably, brings the qubit to lose coherence (right plots). The reason for this is the monotonic increase of the decoherence factor with time (left plots). Three characteristic types of structured environment have been analyzed, namely, (a) Ohmic ($s=1$), (b) sub-Ohmic ($s=0.5$) and (c) super-Ohmic ($s=3$) reservoir.} \label{gammaHrc}
\end{figure}

The behavior of the decoherence factor $\Gamma(T,t)$ is illustrated
in the left panels of Fig.~\ref{gammaHrc} for the three Ohmic regimes 
as a function of the interaction time $t$ and for three different 
values of the reservoir temperature. In the large interaction time 
limit, i.e. when the qubit reaches its stationary 
state, the decoherence factor for both sub-Ohmic and Ohmic 
environments increases with time for all temperatures. This 
means that full decoherence is expected in these cases. On the other 
hand, for the super-Ohmic environment, the decoherence 
factor approaches a constant value, and a residual coherence 
in the qubit state is thus expected.
\par
Motivated by this different behaviour, let us now compare 
the values of the maximum QFI to those of the residual 
coherence \cite{Baumgratz}, as  quantified by the sum of 
the absolute values of the 
off-diagonal elements of the density matrix of the probe 
after the interaction with the reservoir, i.e. 
\begin{equation}
rc=\sum_{j\ne k}\, \big| \varrho_{S,jk}(t)\big| \,.
\end{equation}
In the right panels of Fig.~\ref{gammaHrc}, we show the value 
$H^\text{(opt)}$ of the QFI at the optimal interaction time, together 
with the residual coherence $ rc^\text{(opt)}$ at the same time.  
We notice that in order to achieve higher values of the QFI, 
the probe should interact for enough time with the reservoir and, 
in turn, lose coherence. On the other hand, optimal 
estimation does not necessarily correspond to the strong 
decoherence regime, since in that case no information may be 
left encoded in the qubit state. The optimal conditions for
estimation are thus determined by an interplay  between the 
features of the dephasing dynamics and the specific Ohmic 
structure of the environment, rather than from the sole structure
of the interaction.
\par 
Since this dephasing model is completely solvable at any time, 
it is reasonable to
ask wether quantum non-Markovianity can be considered a resource 
for the estimation of the temperature.
Non-Markovianity of the dephasing map has already been addressed, 
e.g. in Refs.~\cite{addis,haikka13}, where the authors demonstrate 
the  existence of  a critical value 
of the Ohmicity parameter above which memory effects occur, i.e. 
the dynamics is non-Markovian.
This value is temperature-dependent and it is found to vary 
between $s=2$ and $s=3$ moving 
from zero to high temperature. The presence of non-Markovianity 
is witnessed by
a decrease in time of  the decoherence factor $\Gamma(T,t)$ of
 Eq. (\ref{gammaall}). For example, referring to 
Fig. \ref{gammaHrc},  for $s=3$ and $T=0.01$  there exists 
a time interval where $\Gamma$ decreases,
a signature of non-Markovian dynamics. Only in the 
super-Ohmic regime at low 
temperatures memory effects arise. It follows that 
non-Markovianity cannot be linked to 
larger values of the QFI, so it cannot be considered 
a resource for temperature estimation. 
\par
The global estimability of temperature is addressed in Fig. \ref{QSNR}, 
where we show the QSNR evaluated at the optimal interaction time 
$t_\text{opt}$, as obtained from Eqs. \eqref{qsnr} and \eqref{fisher}. We
show the QSNR for the three values of the Ohmicity parameter considered above. 
For low temperatures the QSNR is vanishing,  meaning that in this regime 
the efficiency of any estimation procedure is very low. On the contrary,  for 
higher temperatures it increases. The behavior is $s$-dependent in 
the intermediate temperature regime, whereas for high temperature, \mgap{as already observed above, the structure of the environment becomes irrelevant, and} 
the QSNR  saturates to a universal value, independent of the nature 
of the spectral density of the environment.
\par
\mgap{
We conclude this Section with few more analytic results, valid in 
some specific regimes. 
In the low temperature regime $ T \ll 1 $ and for Ohmic environment $(s=1)$ Eq. (\ref{fisher}) simplifies to
\begin{align}
H (T,t) \stackrel{s=1}{=}\frac{\pi^2 t^2 \left[\pi t T \coth{(\pi t T)}-1\right]^2}{(1+t^2)\sinh^2{(\pi t T)}-\pi^2 t^2 T{^{2}}}\,.
\end{align}
Upon comparing this expression with the numerical results we found
that the above expression provides a good approximation for the 
temperature in the regime $T\lesssim 10^{-2} $. In the high temperature 
limit we obtain instead
\begin{equation}
H (T\gg1,t)= \frac{4(K(t,s)-1)^{2}\bar{\Gamma}[s-2]^{2}}{e^{-4T(1-K(t,s))\bar{\Gamma}[s-2]}-1}\,, \quad{s>2}
\end{equation}
where $ K(t,s)= (1+t^2)^{1-\frac{s}{2}}\cos[(s-2)\arctan(t)] $.
}
\section{\label{sec:5}OPTIMAL MEASUREMENT}
Once the optimal estimation conditions have been determined, a
question naturally arises on whether the corresponding bounds to precision may be achieved in practice, i.e. whether a feasible measurement exists
whose Fisher information is equal to the quantum Fisher information \cite{phaseesti,CMQprob}. 
\par
In order to assess if this happens
in our case, let us consider the most general projective
measurement $\{M_\pm\}$, $M_+ +M_-= {\mathbb I}$ that can be performed 
on the qubit, and let us write the two projectors in the
Bloch representation, i.e. 
\begin{equation}
M_{\pm}=\dfrac{{\mathbb I}\pm\vec{a}\cdot\vec{\sigma}}{2}\,,
\end{equation}
where  $  \vec{a}=[a_{1},a_{2},a_{3}] $, $||\vec{a}||= 1  $ and $\vec{ \sigma} $  consists of three $ 2\times2 $ Pauli matrices, $\vec{\sigma}=[\sigma_{x},\sigma_{y},\sigma_{z}]$. Starting from the most general qubit state (\ref{rho}), the probability 
distribution of the two possibile outcomes is given by 
\begin{equation}
p^{\pm}(t)= \hbox{Tr}[{\varrho_{S}(t)\,M_{\pm}}]=\frac{1}{2}[1 \pm a_{1} e^{-\Gamma}\sin\theta]\,,
\end{equation}
corresponding to a Fisher information
\begin{equation}
F(T)=\sum_{m=\pm}\dfrac{\left[\partial_{T}p^{m}(t)\right]^{2}}{p^{m}(t)}\,.
\end{equation}
The above equation, with the prescription 
$ \theta=\dfrac{\pi}{2} $ for the optimal state, makes it easy to see 
that for a measurement with $ a_{1}=1 $, i.e. the measurement of 
$\sigma_x$ on the qubit, we have $ F(T) = H(T) $ which corresponds 
to optimal estimation of temperature, provided that an efficient estimator
is employed to process the data. Overall, we have that
optimal temperature estimation at the 
quantum limit may be achieved by a feasible strategy, i.e. the 
preparation of the qubit in an eigenstate of $\sigma_x$ and the measurement of the same observable after the interaction with the
thermal sample.
\section{\label{sec:6}CONCLUSIONS}
In this paper we have addressed single-qubit quantum thermometry
by dephasing and we have shown that it provides an effective 
process to estimate the temperature of Ohmic samples. Our scheme 
is inherently quantum, since it exploits the sensitivity of the qubit 
to decoherence, and does not require thermalization with 
the system under investigation.
\par
We found that the QFI has a maximum as a function 
of time at any fixed temperature for Ohmic and sub-Ohmic samples,
whereas in the super-Ohmic case this effect shows up only at high temperatures. 
In turn, this means that optimal estimation of temperature may be achieved 
at a finite interaction time, i.e. before the qubit has reached 
its stationary state. The only case in which the optimal estimation of the temperature
coincides with stationarity and thermalization is for super-Ohmic environments in the low-temperature regime, where a saturation effect is observed.
We also found that in order to achieve higher values of 
the QFI the probe should interact long enough with the sample 
and thus, roughly speaking, lose {\it enough coherence} in order
to gain information about temperature. On the other hand, optimal 
estimation does not necessarily correspond to the strong 
decoherence regime, since in that case no information may be 
left encoded in the qubit state. Our results thus show that 
the optimal conditions for estimation are determined by a non-trivial 
interplay  between the features of the dephasing dynamics 
and the specific Ohmic structure of the environment.
Moreover, we pointed out that non-Markovianity does not play any role
in the temperature estimation protocol, since it is only present in the
dynamics of a probe interacting with a super-Ohmic environment at low temperature,
with no enhancement of the estimation precision.
\par
Finally, we have shown that optimal temperature estimation at the 
quantum limit may be achieved by a feasible strategy, involving preparation of the qubit in an 
eigenstate of $\sigma_x$, observable which is then measured after the interaction with the
thermal sample. Our results also pave the way for future developments,
 including the use of entangled probes and, possibly, a suitable engineering of the interaction Hamiltonian.
 From the perspective of possible realizations of our probing scheme, there are several physical platforms implementing a dephasing dynamics of a two-level system in structured environments. We just mention atomic impurities embedded in Bose-Einstein condensates \cite{HaikkaBEC,CironeBEC} and superconducting qubits \cite{YanSC,VillarSC}.

\section*{Acknowledgements}
This work has been supported by JSPS through FY2017 program (grant
S17118) and by SERB through the VAJRA award (grant VJR/2017/000011).
MGAP is member of GNFM-INdAM and thanks Francesca Gebbia for useful discussions.
\end{document}